\documentclass[multphys,vecphys]{svmult}

\usepackage{makeidx}         
\usepackage{graphicx}        
\usepackage{multicol}        
\usepackage[bottom]{footmisc}
\usepackage{amsmath}
\usepackage{amssymb}

\makeindex             

\begin{document}

\title*{Nuclear activity in galaxies driven by binary supermassive black holes}
\titlerunning{Binary black holes and AGN} 
\author{A.P. Lobanov}
\authorrunning{Lobanov} 
\institute{Max-Planck-Institut f\"ur Radioastronomie, Auf dem H\"ugel 69, 53121 Bonn, Germany.
\texttt{alobanov@mpifr-bonn.mpg.de}}
%
%
\maketitle

\begin{abstract}
Nuclear activity in galaxies is closely connected to galactic mergers
and supermassive black holes (SBH). Galactic mergers perturb
substantially the dynamics of gas and stellar population in the
merging galaxies, and they are expected to lead to formation of
supermassive binary black holes (BBH) in the center of mass of the
galaxies merged. A scheme is proposed here that connects the peak
magnitude of the nuclear activity with evolution of a BBH system. The
scheme predicts correctly the relative fractions of different types of
active galactic nuclei (AGN) and explains the connection between the
galactic type and the strength of the nuclear activity.  It shows that
most powerful AGN should result from mergers with small mass ratios,
while weaker activity is produced in unequal mergers. The
scheme explains also the observed lack of galaxies with two active
nuclei, which is attributed to effective disruption of accretion disks
around the secondary in BBH systems with masses of the primary smaller
than $\sim 10^{10}\,M_\odot$.
\end{abstract}

\section{Binary black holes and nuclear activity in galaxies}

Important roles played by galactic mergers and binary black holes in
galaxy evolution was first recognized several decades ago
\cite{bbr80,roos1981}. A large number of subsequent studies have
addressed the problems of evolution of binary black holes in
post-merger galaxies (see \cite{mm05} for a recent review) and
connection between mergers and nuclear activity in galaxies
\cite{dimatteo03,haehnelt02,springel05}. The correlations observed
between the masses, $M_\mathrm{BH}$, of nuclear black holes in
galaxies, the and masses, $M_\star$, \cite{magorrian1998} and velocity
dispersions, $\sigma_\star$, of the host stellar bulges
\cite{ferrarese2000,gebhardt2000,tremaine2002} suggest a connection
between the formation and evolution of the black holes and galaxies.
Growth of black holes in galactic centers is self-regulated by
outflows generated during periods of supercritical accretion
\cite{fabian1999,haehnelt1998,silk1998}. This mechanism offers a
plausible explanation for the observed $M_\mathrm{BH}$--$\sigma$
relation \cite{king2003}.

Supermassive black holes are expected to form in the early Universe,
with multiple SBH likely to be common in galaxies \cite{haehnelt02}.
However, the detailed connection between the SBH evolution and the
nuclear activity is somewhat elusive. Observational evidence for
binary SBH is largely indirect (see \cite{lobanov2005,mm05} and
references therein), with only two double galactic nuclei (NGC\,6240
\cite{komossa2003} and 3C\,75 \cite{owen1985}) observed directly in
early merger systems, at large separations. There are no convincing
cases for secondary black holes within active galaxies, although some
of them may be hiding among the extranuclear X-ray point sources
detected by ROSAT and Chandra \cite{colbert2002,miller2004}. This
implies that the activity of secondary companions is quenched at early
stages of the merger, possibly due to disruption of the accretion disk.

The nuclear activity depends strongly on the availability of accreting
material in the immediate vicinity of a black hole, and AGN episodes
are believed to last for $\sim 10^7$--$10^8$\,yr. This is likely to be
smaller than typical lifetimes of nuclear binary black holes in
galaxies \cite{mm05}. This suggests that nuclear binary black holes
systems may provide a mechanism necessary for instilling and
supporting high accretion rates over timescales implied by large-scale
relativistic outflows produced in AGN
\cite{dokuchaev91,polnarev94}. Evolutionary stages of BBH systems can
also be connected phenomenologically to different types of AGN
\cite{lobanov2005b}. Here, an analytical model is proposed that
connects the evolution of central SBH in to the nuclear activity in
galaxies.

\subsection{BBH evolution}

The main constituents of the model are: 1)~binary system of
supermassive black holes, 2)~accretion disk, 3)~central stellar bulge.
The BBH is described by the masses $M_1$, $M_2$ ($M_1\ge M_2$) of the
two black hole and their separation $r$.  The accretion disk is
assumed to be a viscous Shakura-Sunyaev disk, with a mass
$M_\mathrm{d}$.  The disk extends from $\rho_\mathrm{in}\,R_\mathrm{g}$ to $\rho_\mathrm{out}\,R_\mathrm{g}$, where $R_\mathrm{g}$ is the
gravitational radius and $\rho_\mathrm{in}\approx 6$ and
$\rho_\mathrm{out} \approx 10^4$ \cite{ivanov+1998}.  The central
bulge extends over a region of radius $r_\star$ and has a mass
$M_\star$ ($M_\star > M_{12} = (M_1 + M_2)$) and a velocity dispersion
$\sigma_\star$. 

The evolution of the BBH is described in terms of \emph{reduced mass},
$\tilde{M}$, and \emph{reduced separation}, $\tilde{r}$ of the binary.
The reduced mass is defined as $\tilde{M} = 2\, M_2\,
/\,M_{12}$.  This definition implies $\tilde{M} = 0$ for $M_2 = 0$ and
$\tilde{M} = 1$ for $M_2 = M_1$. If $q = M_2\,/\,M_1$ is the mass
ratio in the system, then $\tilde{M} = 2\,q \,/(1+q)$.  The reduced
separation is given by $\tilde{r} = r\,/\,(r+r_\mathrm{c})$, where
$r_\mathrm{c}$ is the separation at which the two black holes become
gravitationally bound (this happens at $\tilde{r} = 1/2$). Binary
systems have $\tilde{r}\le 1/2$, while unbound pairs of SBH have
$\tilde{r}>1/2$.

\subsection{Accretion disk disruption in binary black holes}

Two SBH in a merger galaxy are expected to form a binary system at a
separation $r_\mathrm{c} = r_\star\, (M_{12}\, /\,M_\star)^{1/3}$,
with an initial orbital speed $v_\mathrm{init} = \sigma_\star
(M_{12}\,/\,M_\star)^{1/3}$ \cite{bbr80}.  Assuming that the relative
speed of the two black holes reaches asymptotically its Keplerian
value, the \emph{approach speed} can be defined as $v_\mathrm{appr}(r)
= \sigma_\star \,(r_\star\,/\, r)^{1/2}\,(M_{12}\,/\, M_\star)$.  Both
black holes are assumed to have active accretion disks at early
stages of the merger. The separations $r_\mathrm{d}$ at which the
accretion disks are disrupted and eventually destroyed can be
estimated for each of the two black holes by equating the approach
speed to the Keplerian velocity at the outer edge of the disk:
$v_\mathrm{k,out} = c/\sqrt{\rho_\mathrm{out}}$. This yields $r_\mathrm{d} = \rho_\mathrm{out}\, r_\mathrm{c}\,
(\sigma_\star\,/\,c)^{2} \, (M_{12}\,/\,M_\star)^{2} \,$.  The bulge
mass and velocity dispersion must satisfy the
$M_\mathrm{BH}$--$\sigma_\star$ and $M_\mathrm{BH}$--$M_\star$
relations \cite{king2003}. The resulting reduced separation is
$\tilde{r}_\mathrm{d} = 1/(1+\xi)$, where $\xi = M_1 /
[1.86\times
10^7\,M_\odot\, \rho_\mathrm{out}\, \phi^2 \, (2-\tilde{M})^3]$ and $\phi$ is the collimation angle of
the outflow carrying the excess energy and angular momentum from the
immediate vicinity of the black hole. This corresponds to a critical
mass $M_\mathrm{eq} = 1.86\times 10^7\, M_\odot\, \rho_\mathrm{out}\,
\phi^2$ for which an equal mass binary system will undergo disk
destruction at the time of gravitational binding (at
$\tilde{r}_\mathrm{d} = \tilde{r}_\mathrm{c} = 1/2$. In systems with
$M_1 < M_\mathrm{eq}$ the destruction of accretion disk around the
secondary will occur before the formation of a gravitationally bound
system. For typical values of $\rho_\mathrm{out} \approx 10^4$ and
$\phi = 0.1$--$0.3$, $M_\mathrm{eq}$ reaches
$10^9$--$10^{10}\,M_\odot$. It implies that most of active galaxies
formed by galactic mergers should undergo destruction of the disk
around the secondary BH before or during the formation of a
gravitationally bound systems. Since masses of the nuclear black holes
in galaxies rarely exceed $10^{10}\, M_\odot$, this offers a natural
explanation for the observed lack of active galaxies with double
nuclei, since it predicts that \emph{in most galaxies with binary
black hole systems the secondary companion will be inactive}.

\begin{figure}[t]
\centering
\includegraphics[width=0.49\textwidth]{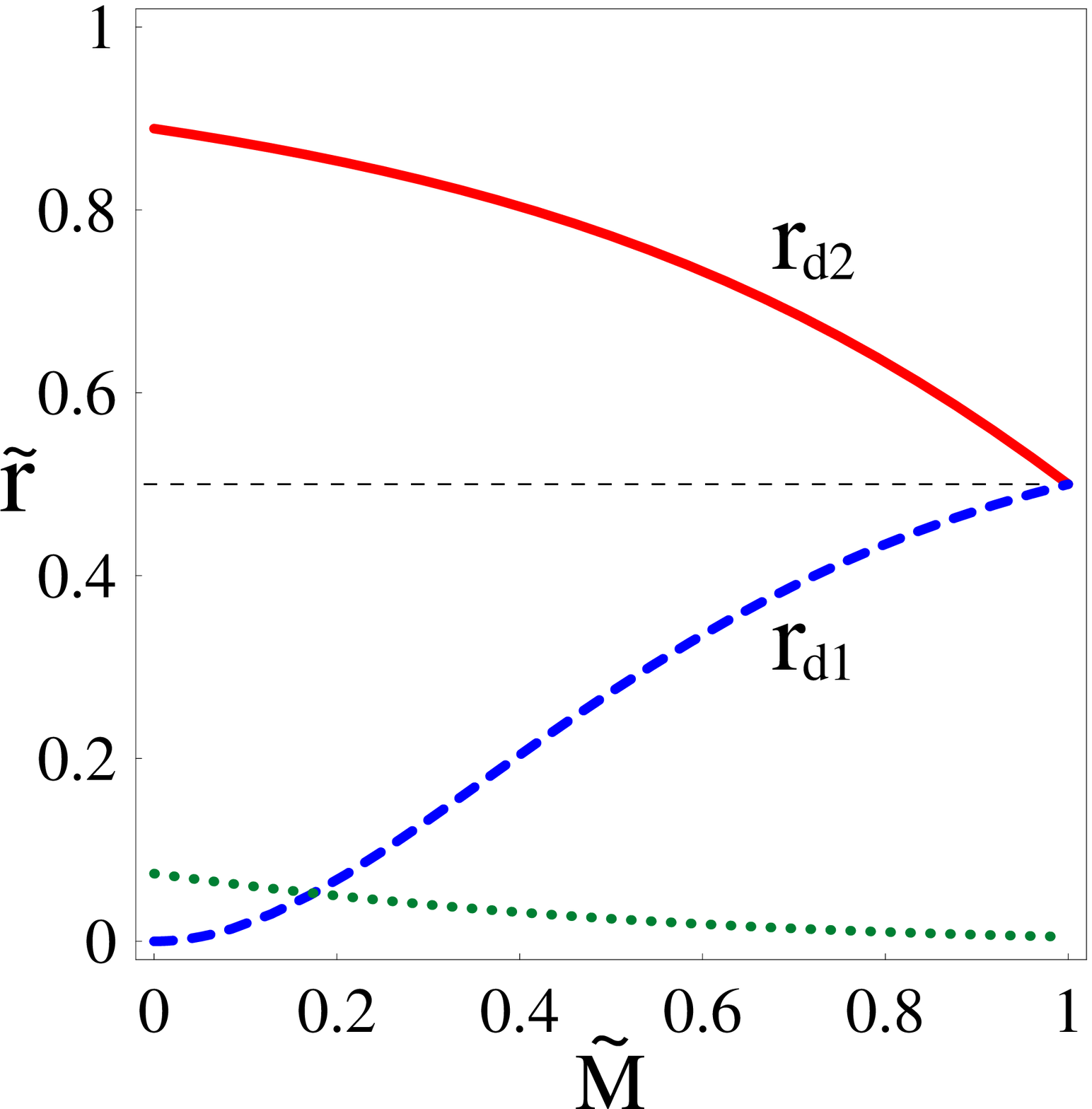}
\includegraphics[width=0.49\textwidth]{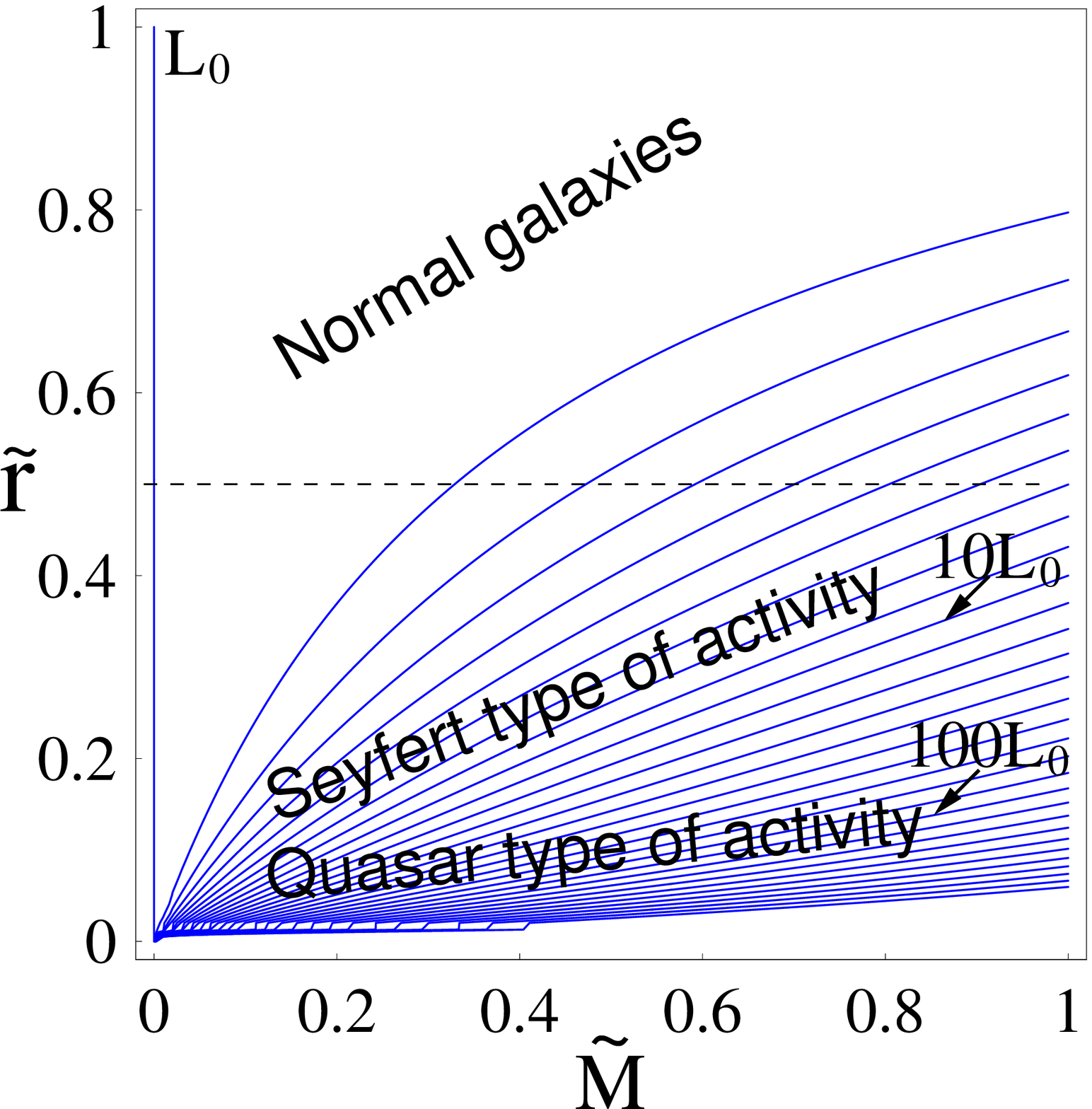}
\caption{Properties of binary black holes in the
$\tilde{M}$--$\tilde{r}$ plane ($\tilde{r}=1/2$ signifies the capture
distance at which a pair of black holes becomes gravitationally
bound). \textbf{Left:}~Reduced separations for the disruption distance
of the accretion disks around the secondary (solid line) and the
primary (dashed line) black holes. The separations are calculated for
$M_1= M_\mathrm{eq}$. Above the $\tilde{r}_\mathrm{d2}$ line, both
black holes retain accretion disks, while only one accretion disk
(around the primary) exists. The dotted line shows the limiting
distance below which a circumbinary disk may exist. This becomes
feasible at $\tilde{M}\lesssim 0.2$. \textbf{Right:}~Peak
luminosities of AGN calculated for a range of values of the reduced
mass $\tilde{M}$ and reduced distance $\tilde{r}$ in binary systems of
SBH in the centers of galaxies. Equal luminosity contours are drawn at
a logarithmic step of 0.1, starting from a unit luminosity $L_0$
marked by the vertical line at $\tilde{M}=0$. }
\label{lobanov1-fig1}
\end{figure}

Denoting $\epsilon_1 = M_1/M_\mathrm{eq}$, the disruption distances are
\[
\tilde{r}_{d1} = \left(1 + \frac{\epsilon_1}{\tilde{M}^2 (2-\tilde{M})}
\right)^{-1}
\quad \mathrm{and} \quad 
\tilde{r}_{d2} = \left(1 + \frac{\epsilon_1}{(2-\tilde{M})^3}
\right)^{-1}\,,
\]
for the primary and secondary black hole, respectively. A circumbinary
disk can exist at orbital separations smaller than $\sim G\, M_1
\rho_\mathrm{out}^{1/2} c^{-2}$. These three characteristic distances
are shown in left panel of Fig.~\ref{lobanov1-fig1}, for $M_1 =
M_\mathrm{eq}$. 

\subsection{Peak luminosity of AGN}

The peak magnitude of the nuclear activity in a galaxy hosting a
binary black hole
system can also be connected with the reduced mass and orbital
separation of the two black holes. Assuming that the accretion rate increases proportionally to the tidal forces acting on stars and gas on scales comparable to the accretion radius, $2\, G\, M_\mathrm{bh}/\sigma_\star^2$,
the peak luminosity from an AGN can be crudely estimated from
\[
L_\mathrm{peak} = L_0 \left( 1 + \frac{\tilde{M}}{2-\tilde{M}}\frac{\tilde{M}}
{\tilde{r}^2}\right)\,,
\]
where $L_0$ is the ``unit'' luminosity of a typical single, inactive
galactic nuclei.  The peak luminosities calculated in this fashion are
plotted in the right panel of Figure~\ref{lobanov1-fig1} for the entire
range of $\tilde{M}$ and $\tilde{r}$.

The peak luminosity
increases rapidly with increasing $\tilde{M}$ and decreasing
$\tilde{r}$, and it reaches $L_\mathrm{peak}=1000\,L_0$ for an equal
mass binary SBH at $r\approx0.03\,r_\mathrm{c}$. This corresponds most
likely to powerful quasars residing in elliptical galaxies. At the
same $\tilde{r}_\mathrm{c}$, an unequal mass binary, with
$\tilde{M}=0.15$, will only produce $L_\mathrm{peak} \approx 10\,L_0$,
which would correspond to a weak, Seyfert-type of active nucleus. Assuming that galaxies are distributed homogeneously in the $\tilde{M}$--$\tilde{r}$ diagram, this
scheme implies that about 70\% of all galaxies should be classified
as inactive, while the Seyfert-type of galaxies, with
$L_\mathrm{peak}=10$--$100\,L_0$, should constitute 25\% of the galaxy
population, and the most powerful AGN, with
$L_\mathrm{peak}>100\,L_0$, should take the remaining 5\%.
It shows that the most powerful AGN
with $L_\mathrm{peak}>1000\,L_0$ should be found in binary SBH with
nearly equal masses of the primary and secondary black holes. Binary
SBH with smaller secondary companions should produce (at the peak of
their nuclear activity) weaker, Seyfert-type AGN. 
Evidence exists in the recent works \cite{laine2003,letawe2006} that the nuclear luminosity does indeed increase with the progression of the merger, but more systematic and detailed studies are required.

\section{Conclusion}

The model described above can be applied effectively to
high-resolution optical studies and data from large surveys that can
be used to obtain estimates of the nuclear luminosities and black
hole masses in active galaxies. The most challenging task is to assess
the state of the putative binary, since the secondary black holes are
very difficult to detect. For wide binaries, Direct evidence may be
sought in galaxies with double nuclei and extranuclear compact
sources. Close binaries can probably be identified only indirectly,
through periodic perturbations caused by the secondary
companion. Other indicators, such as flattening of the galactic
nuclear density profile due to BBH \cite{mm05}, can also be
considered. Once the binary separations have been estimated, it would
be possible to populate the $\tilde{M}$--$\tilde{r}$ diagram and study
whether different galactic and AGN types occupy distinctively different
areas in the diagram.

\end{document}